\documentclass{kluwer}
\usepackage{epsfig}
\begin{document}
\begin{article}
\begin{opening}
\title{The ALFA Laser: Beam Relay and Control System}

\author{T. \surname{Ott}\email{ott@mpe.mpg.de}}
\author{W. \surname{Hackenberg}\email{hacki@mpe.mpg.de}}
\author{S. \surname{Rabien}\email{srabien@mpe.mpg.de}}
\author{A. \surname{Eckart}\email{eckart@mpe.mpg.de}}
\institute{Max-Planck-Institut f\"ur extraterrestrische Physik, Garching}
\author{S. \surname{Hippler}\email{hippler@mpia-hd.mpg.de}}
\institute{Max-Planck-Institut f\"ur Astronomie, Heidelberg}

\begin{abstract}
The ALFA laser subsystem uses a 4 watt continuous wave laser beam to produce an artificial guide star in the mesospheric sodium layer as a reference for wavefront sensing. In this article we describe the system design, focusing on the layout of the beam relay system. It consists of seven mirrors, four of which are motor-controlled in closed loop operation accounting for turbulences inside the dome and flexure of the main telescope.

The control system features several computers which are located close to analysis and control units. The distribution of the tasks and their interaction is presented, as well as the graphical user interface used to operate the complete system. This is followed by a discussion of the aircraft detection system ALIENS. This system shuts off the laser beam when an object passes close to the outgoing laser.
\end{abstract}

\abbreviations{
\abbrev{ALFA}{Adaptive optics with Laser For Astronomy}
\abbrev{AO}{Adaptive Optics}
\abbrev{LGS}{Laser Guide Star}
\abbrev{VME}{Versa Module Europa}
\abbrev{GUI}{Graphical User Interface}
}

\end{opening}

\section{Introduction}

The ALFA project (Adaptive optics with Laser For Astronomy) is a collaboration between the Max-Planck-Institute for Astronomy (MPIA) in Heidelberg which supplies the adaptive optics (wavefront sensing and correction), and the Max-Planck-Institute for extraterrestrial Physics (MPE) who provide the laser guide star (LGS). The complete system is installed on the 3.5~m telescope on Calar Alto, Spain. The adaptive optics subsystem is described in detail in Kasper et al. (1999), and essential aspects of the ALFA Laser are given in Rabien et al. (1999). In this contribution we want to give an overview on the technical system setup and the design of the laser beam relay.

Routine operation of the ALFA system is done by Calar Alto staff. The user interface and software setup has therefore been designed to allow transparent operation without distracting the operator with technical details. Still the possibility remains to fine-tune the system during operation, as well as to gather analysis data for future improvements. The complete system is already available to the astronomical community since mid 1998.

An additional aspect while operating the LGS is to account for airplanes crossing the laser beam. The system has been designed such that under no circumstances a pilot can be affected by the upgoing laser light. For this reason an aircraft detection system has been installed which immediately shuts off the laser whenever an object is located near the laser beam.

\section{Hardware and Software Setup}

The ALFA laser subsystem has been designed to be computer controllable in most of its parts. All the control and analysis algorithms are implemented on industrial type VME-bus machines running the real time operating system VxWorks. For interprocess communication, shared memory techniques are used. The experimental physics and industrial control system EPICS network is used for inter-machine communication using its network database capabilities. This system setup provides reliable operation of the laser while still being very flexible with respect to system hardware changes. The software controlling the aircraft detection system is running on a separate Silicon Graphics Indigo workstation. All software development is done on a SUN workstation which is also used as a file server holding all startup files necessary for booting the VME-bus machines.

\begin{figure}[htp]
\caption{Positions of the beam relay mirrors and general location and tasks of the different computers. All machines are connected using the Calar Alto network system.}
\label{design_overview}
\epsfig{file=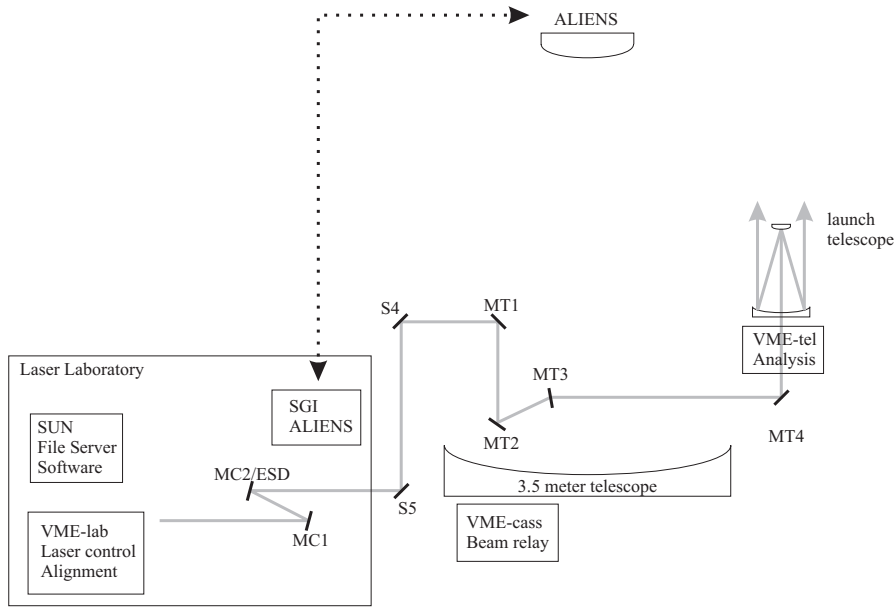, width=\textwidth}
\end{figure}

Figure~\ref{design_overview} shows the locations of the computers and the tasks they are fulfilling. All machines are connected using the Calar Alto network system which is based on twisted-pair cabling. The VME-bus machines have been installed in locations where cable lengths, especially those of the analysis tools, could be dimensioned as short as possible in order to minimise noise. On each of the separate computers, software watchdogs are running which are also visualized in the graphical user interface. In the following sections, we will describe the different tasks and provide information on their operation.

\subsection{Laser Laboratory}

In the laser laboratory on the coud\'e floor of the 3.5m telescope, the machine VME-lab is located. It controls and monitors the two main lasers, their beam properties, and the initial beam injection into the relay system. The tasks running on this computer can be summarized as follows:

\begin{itemize}
\item{Pump laser status monitoring}
\item{Dye Laser frequency adjustment}
\item{Collimation of pilot and dye laser beams}
\item{Control of the polarization state of the dye laser beam}
\item{Initial beam injection into coud\'e path and beam alignment}
\end{itemize}

On the optical bench in the laser laboratory, the beam passes a water-cooled shutter, a quarter-wave plate, and a pre-expander. A beamsplitter and two additional mirrors are used to direct a pilot beam into the path. This beam consists of a small fraction of laser light coupled out from the pump laser and is used for the lower loop which is described in section~\ref{lowerloop}. The two mirrors MC1 and MC2 are used for centering and pointing the dye laser and pilot beams into the relay. The hardware installed on the optical bench for adjusting the collimation of both lasers, the polarization state using a rotatable $\lambda/4$-plate and the initial beam alignment consists of two \textsl{Newport MM~4005} motor controllers which are connected to the VME-bus machine using RS~232C serial lines. Each of these controllers supplies connections to four independent motors.

\begin{figure}[htp]
\caption{Control window of components installed in the laser laboratory.}
\label{laserlab}
\epsfig{file=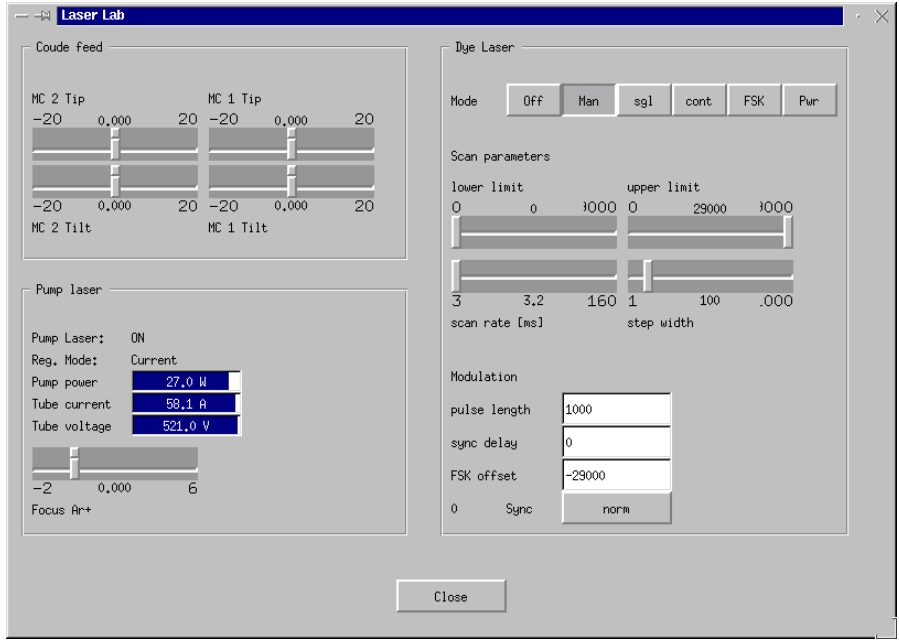, width=\textwidth}
\end{figure}

In Figure~\ref{laserlab} the software control window of the laser laboratory is shown. These components have been put in a separate window to minimise confusion on the main GUI since under normal operating conditions, their values need not be modified. To the lower left, the control parameters of the pump laser are displayed continuously. The focus of the pilot beam can be adjusted to properly focus on the lower loop detectors. On the right hand side, frequency tuning and modulation of the ring dye laser are performed, which are necessary for different experiments. Those are described in Davies et al. (1999) in this volume.

\subsection{Beam Relay}

The relay optics which transport the beam from the coud\'e lab  to the launch telescope are one of the most important parts of the laser system. They are continually undergoing revision, not only to increase their  effectiveness, but also to make them simpler for guest observers to use. Figure~\ref{design_overview} shows the path taken by the laser as it is directed by a succession of actively controlled mirrors. A shorter path with fewer reflections was considered, but involved directing the laser through the centre of the dome where it was not possible to baffle it; it was rejected due to the severe implications for both safety and beam turbulence. To offset the loss from so many components in the path chosen, all mirror surfaces are dielectric broadband coated, and transmissive elements have an  anti-reflection coating optimised to 589~nm. 

The complete relay system is controlled by the machine VME-cass which is mounted on to the Cassegrain flange. The purpose is to keep the beam well aligned in the beam relay system and to reduce beam jitter. A 16 channel analog-digital converter digitizes position signals and feeds a \textsl{Newport PM~500} motion controller with corrective mirror motion commands. 

The relay system consists of a total of six mirrors. The two-axis tracking mirror S5 and S4 on the declination axis are part of the original coud\'e train. Mirror MT1, the declination axis pick-off, marks the end of the lower loop. MT2 is a fixed mirror that directs the beam into the upper loop, consisting of mirrors MT3 and MT4. Before use, the entire relay must be aligned. The initial alignment, which is only needed at the beginning of an observing run, must be done manually. A set of 5 video cameras has been installed at critical points to assist with this. During an observing run, alignment is done automatically by two control loops which are described in the following sections. The control parameters for the loops can be adjusted during operation using a separate control window.

\subsubsection{Lower Loop}
\label{lowerloop}

A "lower guide" control loop is running on VME-cass to keep the laser beam well aligned on the coud\'e axis. Inside the yoke, all degrees of freedom (parallel and angular displacement) are monitored using position sensitive devices (PSD's) which are fed by the pilot laser beam. The analog signals from the PSD's are digitized at a rate of 1~kHz. The difference of those signals to calibrated positions are used in a closed loop operation to move the mirrors ESD and MT1 (which are located in the laser laboratory and on the declination axes, respectively; see Figure \ref{design_overview}) such that proper beam alignment is assured.

The S5 mirror is part of the original coud\'e path of the telescope and has been incorporated in the laser beam relay. It is fully controlled by the telescope control system. During normal operation when the telescope is tracking on a scientific target, tracking errors of the mirror have to be accounted for by the lower guide control loop. In addition, the motion of the coud\'e mirror is not continuous but overlayed with phases of fast motion on timescales of $\sim$5 seconds which have to be corrected as well.

A complication for the lower loop implementation is due to the fact that when slewing the telescope to a new object, the S5 mirror is not fast enough to keep track with the slewing motion of the telescope. Depending on the relative offset, it takes up to one minute until tracking of the S5 mirror proceeds. During this phase, no proper beam alignment is assured and therefore there remains the possibility that the laser beam may leave the beam relay path. Therefore, the status of the S5 mirror is monitored in addition to the total intensity and position of the pilot beam. This information is provided by the telescope control system using an EPICS network variable.

When either the intensity drops below normal values, the position signal is lost or the status of the S5 mirror indicates slewing of the telescope, the dye laser beam is shut off in the laser laboratory. When slewing of the S5 mirror is finished, a scanning phase begins to automatically realign the pilot beam. Only when the intensity and position signals indicate proper alignment is the shutter of the dye laser opened again.

\subsubsection{Upper Loop}

Once the laser beam is picked off the declination axes, it enters the "upper loop". This loop provides the final centering and alignment of the beam in the relay. It runs at high frequencies ($\sim$100~Hz) to be able to correct for beam jitter introduced by turbulent air along the relay. The beam position and angular displacement are measured directly below the launch telescope using a PSD and a triple-prism detector which sample the dye laser itself. Just like the "lower loop", the digitized values of the detectors are used to calculate correction commands for the two fast steering mirrors MT3 and MT4.

When the 3.5~m telescope is slewed to positions at extremely low altitudes or hour angles, flexure of the main telescope body becomes an important issue. Even when properly aligned on the coud\'e path, the beam may not hit the secondary of the launch telescope perfectly. To account for these extreme telescope positions, a scanning phase similar to the one of the lower loop is planned to be implemented in the near future.

\subsubsection{Baffling}

The laser light passes through the dome from the exit of the laser laboratory up to the main body of the telescope. Temperature differences of several degrees are common between the surrounding air and the main body of the telescope. This results in turbulent air which distorts laser beam quality. In order to enhance beam quality, most of the beam relay system has therefore been baffled using black anodized aluminum tubes. This also serves as a safety precaution to prevent persons from looking directly into the collimated beam.

\begin{figure}[htp]
\caption{The coud\'e mirror S5 with the baffles coming from the laser lab (bottom) and going to the telescope yoke (right).}
\label{s5}
\epsfig{file=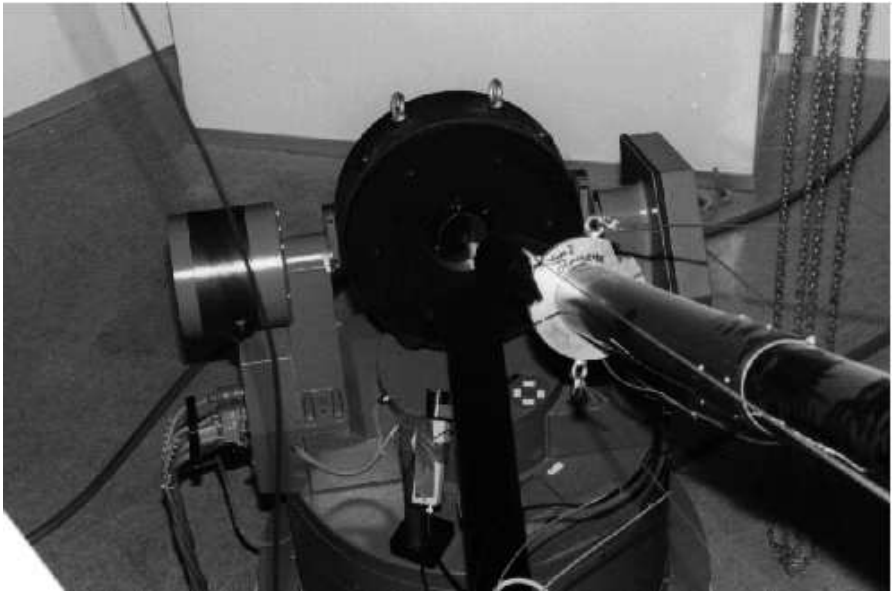, width=\textwidth}
\end{figure}

In Figure~\ref{s5} we show a photograph of the first beam relay mirror S5 with the baffles coming from the laser laboratory and going to the main body of the telescope into the yoke. The baffle coming from the laser lab is closed by a window on the optical bench in the laser laboratory to prevent turbulent air streaming from the warmer laser lab to the S5 mirror due to a chimney effect. This mirror and its support itself are also planned to be covered completely.

\subsection{Launch Telescope}

The launch telescope (50~cm primary and 5~cm secondary mirrors) is situated 2.9~m off-axis from the main telescope primary. It is an afocal Galilean-type beam expander through which the laser can be projected both on- and off-axis. The main disadvantage of launching on-axis is the power lost due to the obscuration by the secondary mirror. The power loss is 1.5--3\% depending on the diameter of the exit beam, which can be varied in the range 24--49~cm by adjusting the pre-expander. More significant power loss is from the edges when the beam diameter is similar to that of the primary, but this does not affect the central intensity of the resulting LGS which is the important parameter for good adaptive optics correction.

The secondary mirror of the launch telescope is mounted on a piezo platform used for pointing the laser at science targets. It has a full field of view of 50'', with a positional accuracy of $\sim$0.05'' and a maximum steering frequency of 30~Hz. Focussing the laser in the mesosphere is done using a stepper motor with a resolution of $\sim$500~m in height. In the near future, a link from the AO bench will be installed providing LGS tip/tilt information so that the mirror can be used to compensate for beam jitter introduced in the atmosphere.

The VME-tel machine is located in a rack next to a analysis breadboard mounted just below the entrance of the launch telescope. On this breadboard, several analysis tools are installed which are either controlled via serial lines or whose analog outputs are digitized by a 16 channel 12 bit analog-digital converter. In addition, the secondary mirror of the launch telescope is controlled. The different tools are:

\begin{itemize}
\item{Offline: Power measurement}
\item{Offline: Beam collimation}
\item{Polarization status}
\item{Beam wavefront}
\item{Tip Tilt of secondary}
\item{Focus of launch telescope}
\end{itemize}

Immediately before being projected into the atmosphere, the quality of the beam can be checked by a diagnostics bench. The offline analysis tools are inserted using a linear stage which moves mirrors to reflect the laser light onto them. The linear stage is controlled by a \textsl{Newport PM~500} motion controller. The polarization state is measured online with a division-of-amplitude device mounted near the secondary mirror of the launch telescope. On the specific layout of the analysis tools and measurement results, see Rabien et al. (1999) in this volume.

\subsection{Graphical User Interface}

The complete laser system is controlled with a graphical user interface (GUI). The GUI runs on the SUN workstation located in the laser lab and can be displayed on any workstation or X-Terminal on Calar Alto. During an observing run, control is performed using a separate X-Terminal in the control room of the telescope.

\begin{figure}[htp]
\caption{The graphical user interface used to control the laser system.}
\label{laserleitstand}
\epsfig{file=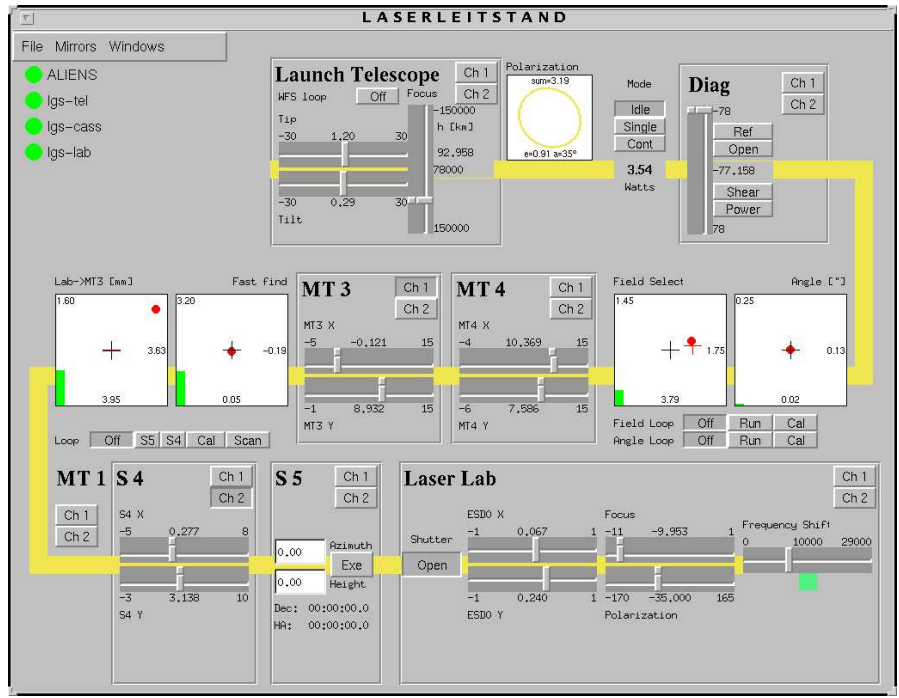, width=\textwidth}
\end{figure}

The GUI (Figure~\ref{laserleitstand}) has been designed such as to keep the abstraction level as low as possible in order to perform transparent operation. The several components of the laser system are visualized, from the laser laboratory to the lower right to the launch telescope at the upper left. All mirrors and other motor controlled devices can be moved using sliders. As a result of any slider movement, an EPICS variable is set which is monitored on the VME-bus machines which then command the motor movements.

The two control loops and the signals of the beam position measurement devices are visualized in the central part of the GUI. Control parameters for the loops can be adjusted using a separate window, which is only necessary when reconfiguring the laser system or at the beginning of an observing run. The polarization status is displayed on the GUI to the right of the launch telescope, as well as the offline power measurement.

The signals of the various CCD cameras that have been installed along the beam relay path are displayed on a video monitor which is located in the control room next to the X-Terminal. In the GUI, it is possible to select which camera signal to be displayed using the buttons with the camera symbol. This enables the operator to check the beam position on the separate mirrors for beam alignment purposes.

In the upper left part, the status of the different computers is displayed using their watchdog programs. This way the user is able to check the overall status of the system, enabling him to interfere in case of system failures.

\section{Aircraft Detection}

Of particular concern for the safe operation of a LGS adaptive-optics
system is the protection of aircraft passing the upgoing laser-beam. Although
the probability of hitting not only the aircraft, but also having the pilot look directly into the beam is extremely low, precautions have to be taken to avoid this situation because even an expanded 4 Watt continuous wave laser could affect and distract the pilot for a few minutes. The scattered light of such a laser is also too dim to allow the pilot to see it before entering the beam.

Besides reliably detecting objects a useful system is also expected to generate only a very limited amount of false alarms, since every alarm immediately shuts down the laser beam and with that many minutes of integration time of the scientific instrument may be wasted. The requirements for the Calar Alto system are a rate of false alarms of less than 0.1 per hour. A spotter is not required at Calar Alto since the output power of the outgoing laser beam is too low, as opposed to pulsed systems with peak powers of up to 1~kW.

\subsection{Visible CCD Detectors}

All aircraft are required to use position and anti collision lights (see Federal Aviation Requirements). The anti collision light is a yellow, red or white light, flashing with a frequency of 40 to 100 flashes per minute. The position lights are red and green lights, located at the tips of the wings, and a white light at the tail. The detection system uses a CCD camera and an image processing system to detect this radiation.

Particularly difficult situations occur when the aircraft is passing the observatory at a low altitude. The time from entering the field of view of the camera to crossing the laser beam may be as low as 2 seconds and therefore the reaction time of the system should not be greater than 1 second. The frequency of the anti collision light may be too low for reliable detection, however, at these low altitudes, the position lights are easily visible. These CCD systems can also easily detect some satellites at low altitudes.

\subsection{The Aircraft Light Imaging Emergency Notification System (ALIENS)}

For the ALFA project, it was decided to mount the CCD camera behind the secondary mirror of the main telescope. ALIENS was developed by Adaptive Optics Associates, Inc. in Boston, MA for the use on the ALFA system and is integrated into the ALFA software in a failsafe way. It is using the EPICS realtime-database to communicate with ALFA and the telescope control software. The specifications for ALIENS are:

\begin{itemize}
\item{Aircraft viewing angle: $< 75^{\circ}$ from horizontal plane}
\item{Range: 0.5 to 30~km}
\item{Aircraft Velocity: $< 250$~m/s}
\item{Operating conditions: Night, Clear Sky}
\item{Environmental Conditions: Temperature $-10^{\circ}$C to $50^{\circ}$C, Humidity $< 90\%$}
\item{Field of View: $10^{\circ}$ Radius}
\item{Rate of False Alarms: $< 0.1$ per hour}
\item{Wavelength Range: 600-1000nm}
\end{itemize}

There are various ways how to detect moving objects in the sky. The approach taken by ALIENS is to watch the sky with a CCD camera and look at variations between consecutive frames.

\subsubsection{Hardware}

The images analysed by ALIENS are taken with a CCD camera which is mounted on the front ring of the telescope behind the secondary mirror. The CCD chip has a total of 640x480 pixels. The field of view of the camera was chosen to be 20 degrees which is approximately the same field of view as seen through the dome slit from behind the secondary.

\begin{figure}[htp]
\caption{Block diagram of the ALIENS aircraft detection system. Behind the secondary of the 3.5m telescope, the CCD camera with its control hardware is located. Via an optical fibre link, the video signals as well as the serial control connection are connected to a Silicon Graphics workstation located in the laser lab. There the control software and aircraft detection algorithm are running. On positive detection, an EPICS variable is set which causes the VME-bus computer which controls the shutter to send a signal so the shutter is closed.}
\label{aliens_block}
\epsfig{file=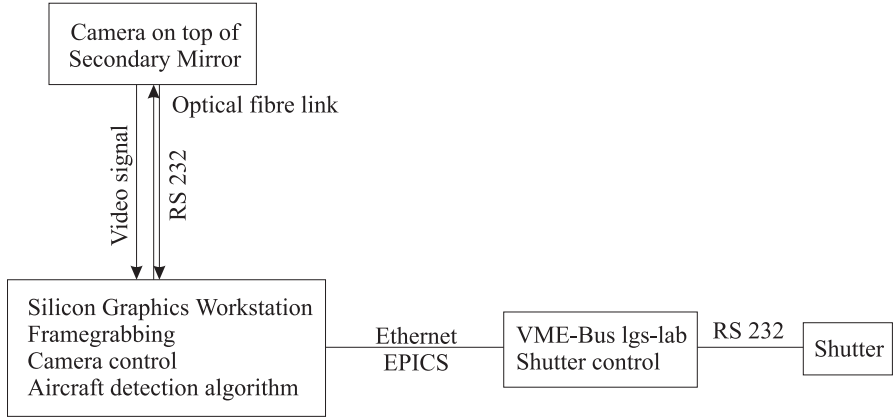, width=\textwidth}
\end{figure}

The single images are transfered by an optical fiber link from the secondary mirror cell to a Silicon Graphics workstation which is equipped with a video capture board. There the images are digitized and further analysed. In Figure~\ref{aliens_block} a block diagram of the hardware used in the ALIENS system is shown.

\subsubsection{Algorithm for aircraft detection}

The most simple approach to detect moving objects such as airplanes
on optical images is to use their relative motion against the stars.
To do this, two consecutive frames received from the CCD camera are
subtracted from each other. As the 3.5m telescope on Calar Alto is
mounted equatorially, objects like stars should cancel out perfectly
when subtracting two frames since there is no field rotation. So whenever
there is a significant difference between two frames, this can be
taken as a possible detection of an aircraft in the field of view.

There are, however, some problems which forbid this simple approach.
For instance, slowly moving objects like planets can trigger an alarm since they are very bright. Even stars can trigger alarms due to scintillation as the transparency of the atmosphere is not constant as a function of time and causes the stars to flicker.

It is therefore necessary to eliminate the signal of bright steady
objects in a different way. This can be done by calculating a mask from
a single image in which all bright objects are marked. When calculating
the difference of two frames, objects that fall into the masked regions are neglected.
This procedure has proven to nearly eliminate all false alarms due to
stars. This mask can be recalculated after a specific amount of time
has elapsed to account for planets, which move slowly but significantly
during an observing night.

A second, fixed mask can be used in addition to the one just described.
This is useful to account for the different motions of the telescope
itself and the dome. When the dome is starting to block a portion
of the field of view seen by the CCD camera, faint stars in this area which are not included in the first mask
will vanish and thus trigger a false alarm. Secondly, the Rayleigh
backscatter of the projected laser beam is usually variable and should
be disregarded in the aircraft detection algorithm.

When accounting for both masks in the image subtraction process, the final
criterion for an aircraft detection is that the difference of the
two frames exceeds a certain threshold. This threshold is just slightly
above the inherent read noise of the CCD chip to be as sensitive as
possible. When an airplane has been detected, several actions are
taken:

First and most important, the shutter is closed by a software command. Since alarms will be generated as long as the airplane is in the field of view of the camera, the shutter will be opened again after a specified time following the last aircraft alarm. In addition, a warning program is started to visually inform the observing astronomer that the LGS is not available. Optionally, several frames can be saved in order to analyse the data and identify eventually false alarms. This is done to further improve the system and its reliability.

\begin{figure}[htp]
\caption{The ALIENS graphical user interface. In the most part of the window, the frames taken by the CCD camera are displayed. The buttons on the bottom left are used to start and stop the system, to pop up a window which controls the software parameters, to display a brief description of the usage and to quit the application.}
\label{awatch}
\epsfig{file=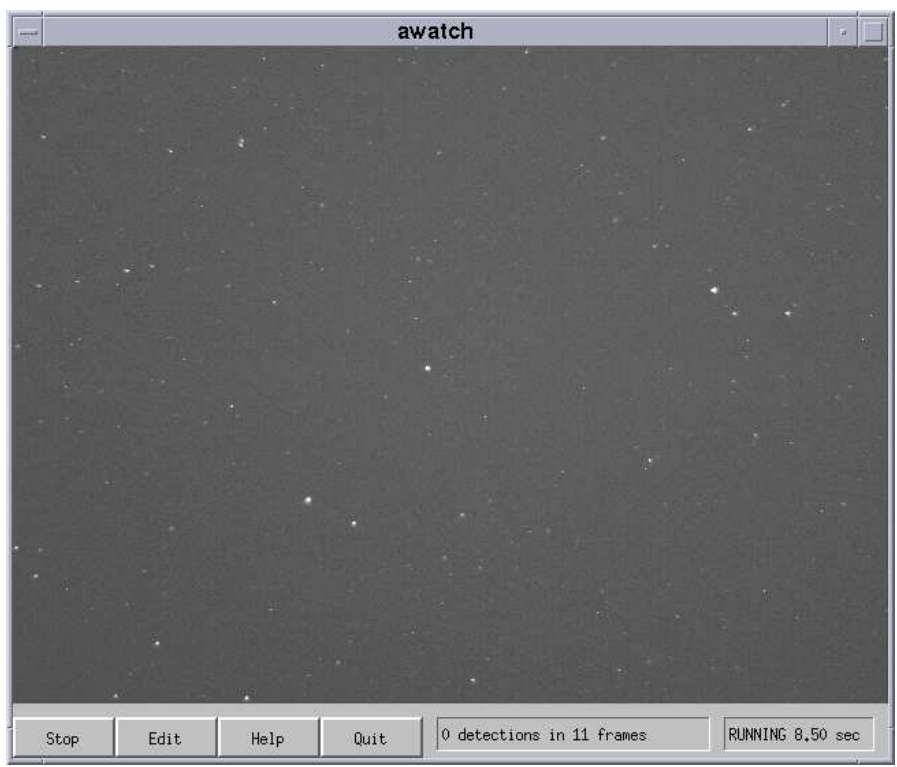, width=\textwidth}
\end{figure}

A graphical user interface has been developed to control the system parameters and visually inspect the online data from the camera on a computer monitor (see Figure~\ref{awatch}). With the computer's X-Window system, this user interface allows network transparent operation of the system with the data processing taking part on one of the workstations in the laser lab and a control of the system from one of the X-terminals in the telescope's control room. On the main user interface for the laser control, the status of ALIENS is displayed as well as a visual reminder for the laser operator.

\subsection{Operation}

During the acceptance test near Boston, MA, the system was operated
for 5 hours without generating any false alarm. With the same settings
of parameters and looking into a different direction it was able to reliably
detect all of the visible aircraft. This included low altitude planes
approaching Boston airport, as well as high altitude planes with only
their anti collision lights visible.

Since March 1998, the system is in full operation whenever the LGS is used on Calar Alto. It proved very reliable during normal observing conditions with less than one false alarm per observing night. The false alarms are predominantly:

\begin{itemize}
\item{Clouds being illuminated by the moon}
\item{Shooting stars crossing the field of view}
\item{Beam alignment problems}
\end{itemize}

Shooting stars cause the laser to be shut off for about 1 second since they cross the field of view very rapidly. False alarms due to clouds cause no major problem to the operation of the laser since under cloudy conditions, usage of the LGS is not possible due to the faint brightness of the laser beacon. The laser shutter is closed whenever there are problems due to beam alignment. This is caused by the raised background in the images because the laser does not shoot into the night sky but is blocked by the telescope dome. When doing the initial beam alignment in closed dome, it is therefore necessary to stop ALIENS in order to be able to visually inspect beam alignment on the telescope dome.

\end{article}
\end{document}